\documentclass[secnumarabi,superscriptaddress,amssymb,nobibnotes,aps]{revtex4}

\usepackage{epsfig}
\usepackage{graphicx}
\usepackage{amsmath}
\usepackage{hyperref}

\begin{document} 

\title{Nucleon dissociation and incoherent $J/\psi$ photoproduction on nuclei in ion ultraperipheral collisions at the CERN Large Hadron
Collider}

\author{V. Guzey}

\affiliation{Department  of Physics,  University  of  Jyv\"askyl\"a, P.O. Box 35, 40014  University  of  Jyv\"askyl\"a,  Finland} 
\affiliation{Helsinki Institute of Physics, P.O.  Box  64,  00014  University  of  Helsinki,  Finland} 
\affiliation{National Research Center ``Kurchatov Institute'', Petersburg Nuclear Physics Institute (PNPI), Gatchina, 188300, Russia}

\author{M. Strikman}
\affiliation{Department of Physics, The Pennsylvania State University, State College, PA 16802, USA}

\author{M. Zhalov}
\affiliation{National Research Center ``Kurchatov Institute'', Petersburg Nuclear Physics Institute (PNPI), Gatchina, 188300, Russia}


\begin{abstract} 

Using the general notion of cross section fluctuations in hadron--nucleus scattering at high energies, we derive an expression for 
the cross section of incoherent $J/\psi$ photoproduction on heavy nuclei $d\sigma_{\gamma A \to J/\psi Y}/dt$, which includes both elastic $d\sigma_{\gamma p \to J/\psi p}/dt$ and proton-dissociation $d\sigma_{\gamma p \to J/\psi Y}/dt$ photoproduction on 
target nucleons. 
We find that with good accuracy, $d\sigma_{\gamma A \to J/\psi Y}/dt$ can be expressed  as a product of the sum of the 
$d\sigma_{\gamma p \to J/\psi p}/dt$ and $d\sigma_{\gamma p \to J/\psi Y}/dt$ cross sections, which have been measured 
at HERA, and
the common nuclear shadowing factor, which is calculated using the leading twist nuclear shadowing model. 
Our prediction 
for the cross section of incoherent $J/\psi$ photoproduction in Pb-Pb UPCs at $\sqrt{s_{NN}}=2.76$ TeV
and $y=0$, $d\sigma_{AA \to J/\psi AY}(y=0)/dy=0.59-1.24$ mb, agrees within significant theoretical uncertainties with the ALICE data.

\end{abstract}

\maketitle

\section{Introduction}
\label{sec:intro}
 
It is now widely accepted that ultraperipheral collisions (UPCs) of relativistic ions, which are characterized by scattering 
at large impact parameters such that the interaction proceeds via emission of quasi-real photons, offers unique possibilities to explore photon--photon and photon--hadron (proton, nucleus) interactions at previously unattainable high 
energies~\cite{Baltz:2007kq}. Focusing on UPC studies of nuclear structure in QCD at the Large Hadron Collider (LHC), 
for a recent summary, 
see, e.g.~\cite{Klein:2017vua}, new constraints on the small-$x$ gluon density in heavy nuclei have been obtained using
coherent $J/\psi$ photoproduction in Pb-Pb UPCs at $\sqrt{s_{NN}}=2.76$ TeV~\cite{Abbas:2013oua,Abelev:2012ba,Khachatryan:2016qhq}. 
The analysis~\cite{Guzey:2013xba,Guzey:2013qza} of these data showed that they give first direct and essentially model-independent
evidence of large nuclear gluon shadowing, $R_g=g_A(x,\mu^2)/[A g_N(x,\mu^2)] \approx 0.6$ at $x=10^{-3}$ and $\mu^2=3$ GeV$^2$
($g_A$ and $g_N$ are gluon densities in Pb and the proton, respectively).
This is consistent with predictions of the leading twist nuclear shadowing model~\cite{Frankfurt:2011cs}, which have small theoretical uncertainties is this kinematics, and also agrees with the EPS09~\cite{Eskola:2009uj}, EPPS16~\cite{Eskola:2016oht}, and nCTEQ15~\cite{Kovarik:2015cma} nuclear parton distribution functions (nPDFs), which however have significant uncertainties in this kinematic region.

At the same time, predictions of the leading twist nuclear shadowing model significantly underestimate~\cite{Guzey:2013jaa} the cross section of incoherent $J/\psi$ photoproduction in Pb-Pb UPCs at $\sqrt{s_{NN}}=2.76$ TeV and $y=0$~\cite{Abbas:2013oua}. It was hypothesized in Ref.~\cite{Guzey:2013jaa} that the process of $J/\psi$ photoproduction with nucleon dissociation $\gamma N \to J/\psi Y$, which was not included in that analysis, may contribute to incoherent $J/\psi$ photoproduction on nuclei and improve the agreement between theory
and experiment. In this article, we demonstrate that this is indeed the case. 
Extending the formalism of Ref.~\cite{Guzey:2013jaa} to include the $\gamma N \to J/\psi Y$ contribution to the nuclear incoherent 
cross section
and using the HERA data on the elastic and proton-dissociative $J/\psi$ photoproduction cross sections~\cite{Alexa:2013xxa}, we show that 
the predictions of the leading twist nuclear shadowing model for incoherent $J/\psi$ photoproduction in Pb-Pb UPCs at $\sqrt{s_{NN}}=2.76$ TeV and $y=0$ agree with the available ALICE measurement.
  
 The rest of the paper is organized as follows. In Sec.~\ref{sec:fluct}, we introduce nucleon cross section fluctuations and their connection to the elastic and proton-dissociation $J/\psi$ photoproduction cross sections.
 The derivation of the incoherent cross section of $J/\psi$ photoproduction on nuclear targets using the  leading 
 twist nuclear shadowing model and generic representations of hadronic fluctuations of the projectile photon 
 and target nucleons is given in Sec.~\ref{sec:cs}. The application of this result to Pb-Pb UPCs in the LHC kinematics and comparison to the ALICE data are presented in Sec.~\ref{sec:upc}. We draw our conclusions in Sec.~\ref{sec:conclusion}.

\section{Cross section fluctuations and proton-dissociation $J/\psi$ photoproduction}
\label{sec:fluct}

At high energies, diffractive dissociation can be understood in the Good--Walker picture in terms of coherent hadronic fluctuations, 
which diagonalize the scattering operator~\cite{Good:1960ba}. Applying these ideas to elastic and proton-dissociation $J/\psi$ photoproduction, one obtains for the respective cross sections at the momentum transfer $t=0$:
\begin{eqnarray}
\frac{d\sigma_{\gamma p \to J/\psi p}(W_{\gamma p},t=0)}{dt} &=& \frac{\varkappa^2}{16 \pi} 
\left(\sum_i |c_i|^2  \sigma_i\right)^2 \equiv \frac{\varkappa^2}{16 \pi} \langle \sigma \rangle^2 \,, \nonumber\\
\frac{d\sigma_{\gamma p \to J/\psi Y}(W_{\gamma p},t=0)}{dt} &=& 
\frac{\varkappa^2}{16 \pi}  \left[\sum_i
|c_i|^2 \sigma_i^2- \left(\sum_i |c_i|^2 \sigma_i\right)^2 \right]=
\frac{\varkappa^2}{16 \pi}  \left[\langle \sigma^2 \rangle- 
\langle \sigma \rangle^2 \right]
 =\frac{\varkappa^2}{16 \pi} \omega_{\sigma}  \langle \sigma \rangle^2 \,,
\label{eq:fl1}
\end{eqnarray}
where $\varkappa$ is proportional to the $\gamma - J/\psi$ transition amplitude; 
$|c_i|^2$ and $\sigma_i$ are the probability and the corresponding
eigenvalue (cross section) for a given fluctuation to contribute to the scattering cross section. 
Note that the index $i$ can also be continuous and multidimensional.
The second of Eqs.~(\ref{eq:fl1}) demonstrates that diffractive dissociation is possible only if various fluctuations interact with different $\sigma_i$, i.e., the distribution over the fluctuations has a non-vanishing dispersion 
$\omega_{\sigma}=\langle \sigma^2 \rangle/\langle \sigma \rangle^2-1$~\cite{Miettinen:1978jb}.

Equation~(\ref{eq:fl1}) is general and admits different interpretations in terms of microscopic models of the nucleon structure.
In particular, it was interpreted in terms of fluctuations of the gluon density in the proton
in Ref.~\cite{Frankfurt:2008vi} 
and fluctuations of the proton shape in 
Refs.~\cite{Mantysaari:2016ykx,Mantysaari:2016jaz,Cepila:2016uku,Traini:2018hxd}.
Note also that the latter in the context of the chiral magnetic effects were considered in~\cite{Kharzeev:2017uym};
the influence of proton size fluctuations on the number of wounded nucleons was studied in~\cite{Alvioli:2013vk,Alvioli:2014sba}.

Extension of Eq.~(\ref{eq:fl1}) to $t \neq 0$ requires the assumption  
that the fluctuations do not mix and a specific model for the distribution of fluctuations in the transverse (impact parameter $\vec{b}$) plane, see, e.g.~Ref.~\cite{Miettinen:1978jb}. 
Indeed, the formalism of cross section fluctuations, which leads to Eq.~(\ref{eq:fl1}), is applicable only 
for very small $|t| \ll 1/R_T^2$  ($R_T$ is the target size); 
for larger $t$, the coherence among the eigenstates $i$ implied in Eq.~(\ref{eq:fl1}) is lost~\cite{Frankfurt:1994hf}
and the dynamics of diffraction dissociation changes. In particular, in the limit of large $|t| > 2$ GeV$^2$, 
the $\gamma p \to J/\psi Y$ process may proceed via the perturbative mechanism of the two-gluon exchange~\cite{Frankfurt:2008et}.
In the following, we assume that the scattering amplitudes corresponding to the elastic and proton-dissociation final states
have the following forms 
in impact parameter space, respectively:
\begin{eqnarray}
\Gamma_{N}(\vec{b}) &=& \frac{\langle \sigma \rangle }{4 \pi B_{\rm el}} e^{-\vec{b}^2/(2B_{\rm el})} \,, \nonumber\\
\Gamma_{Y}(\vec{b}) &=& \frac{\sqrt{\omega_{\sigma}}\langle \sigma \rangle }{8 \pi^2} \int d^2 q_t^{\prime} \,e^{-i \vec{q^{\prime}_t} \vec{b}} f_{\rm pd}(t^{\prime}) \,,
\label{eq:Gamma_N}
\end{eqnarray}
where $B_{\rm el}$ is the slope of the elastic cross section; $f_{\rm pd}(t)$ parametrizes the $t$ dependence of the proton-dissociation cross section (see Eq.~(\ref{eq:tdep}) below).
Using Eq.~(\ref{eq:Gamma_N}), Eq.~(\ref{eq:fl1}) can be generalized to the $t \neq 0$ case as follows:
\begin{eqnarray}
\frac{d\sigma_{\gamma p \to J/\psi p}(W_{\gamma p},t)}{dt} &=& \frac{\varkappa^2}{4 \pi} 
\left|\int d^2 b\, e^{i \vec{q}_t \vec{b}}\, \Gamma_{N}(\vec{b})\right|^2 =\frac{\varkappa^2}{16 \pi} \langle \sigma \rangle ^2 e^{-q_t^2 B_{\rm el}} \,, \nonumber\\
\frac{d\sigma_{\gamma p \to J/\psi Y}(W_{\gamma p},t)}{dt} &=& \frac{\varkappa^2}{4 \pi} 
\left|\int d^2 b\, e^{i \vec{q}_t \vec{b}} \Gamma_{Y}(\vec{b})\right|^2
= \frac{\varkappa^2}{16 \pi} \omega_{\sigma} \langle \sigma \rangle^2 [f_{\rm pd}(t)]^2 \,,
\label{eq:tdep}
\end{eqnarray}
where $t=-q_t^2$.
Based on our discussion above, Eq.~(\ref{eq:tdep}) can be viewed as an interpolation between the $t=0$ and large $|t|$ 
regimes, whose parameters are determined by available data. 

In our analysis, we do not employ a particular dynamical realization for the probabilities $|c_i|^2$  in Eqs.~(\ref{eq:fl1}), (\ref{eq:Gamma_N}), and (\ref{eq:tdep}) and use instead the H1 data on elastic
and proton-dissociation $J/\psi$ photoproduction~\cite{Alexa:2013xxa}. Using these data, we find that 
$\omega_{\sigma}=0.29 \pm 0.04$ for $m_p < M_Y < 10$ GeV and $40 < W_{\gamma p} < 110$ GeV (these values of $W_{\gamma p}$ overlap 
with those probed in Pb-Pb UPCs at $\sqrt{s_{NN}}=2.76$ at central rapidities).
Further, for the measured ratio of the $t$ integrated cross sections, we find $r=\sigma_{\gamma p \to J/\psi Y}(W_{\gamma p})/\sigma_{\gamma p \to J/\psi p}(W_{\gamma p})=0.83 \pm 0.15$ for $\langle W_{\gamma p} \rangle =93.3$ GeV; note also that
$r$ is a slow function of $W_{\gamma p}$ on the studied interval of $W_{\gamma p}$. 
The large value of $r$ is a consequence of the fact that the $t$ dependence of 
 $d\sigma_{\gamma p \to J/\psi Y}/dt$ is much slower than that of $d\sigma_{\gamma p \to J/\psi p}/dt$.
As we show in our numerical analysis in Sec.~\ref{sec:upc}, it is the large cross section of proton-dissociation $J/\psi$ photoproduction on the nucleon (the large value of $r$), which increases our theoretical predictions for the cross
section of incoherent $J/\psi$ photoproduction on nuclei by almost a factor of two and brings it in agreement with the ALICE data.

\section{The cross section of incoherent $J/\psi$ photoproduction on nuclei}
\label{sec:cs}

To include the $\gamma N \to J/\psi Y$ contribution to the nuclear incoherent cross section, we extend our analysis in~\cite{Guzey:2013jaa} by taking into account the effect of cross section fluctuations discussed in Sec.~\ref{sec:fluct}. At high energies, the incoherent $\gamma A \to J/\psi Y$ cross section can be written in the
following form (the nuclear final state $Y$ contains products of nucleus dissociation $A^{\prime}$ caused by the elastic $\gamma N \to J/\psi N$ and inelastic $\gamma N \to J/\psi Y$
processes on target nucleons):
\begin{eqnarray}
\frac{d \sigma_{\gamma A \to J/\psi Y}(W_{\gamma p})}{dt}
&=& \frac{\varkappa^2}{4 \pi} \sum_{A^{\prime}}  \left|\int d^2 b\, e^{i \vec{q}_t \vec{b}} 
\langle A^{\prime}|\Gamma_A(b)|0 \rangle \right|^2-\frac{d \sigma_{\gamma A \to J/\psi A}}{dt}\nonumber\\
&=&\frac{\varkappa^2}{4 \pi} \int d^2 b  \int d^2 b^{\prime} e^{i \vec{q}_t (\vec{b}-\vec{b}^{\prime})} 
\left[\langle 0|\Gamma_A^{\dagger}(b^{\prime})\Gamma_A(b)|0 \rangle -
\langle 0|\Gamma_A^{\dagger}(b^{\prime})|0 \rangle \langle 0|\Gamma_A(b)|0 \rangle \right]
\,,
\label{eq:cs1}
\end{eqnarray}
where $\langle 0|\dots|0 \rangle$ denotes averaging over the nuclear ground-state wave function;
$\Gamma_A$ is the nuclear amplitude;
$d \sigma_{\gamma A \to J/\psi A}/dt$ is the coherent nuclear cross section.
In the last line of Eq.~(\ref{eq:cs1}), we used the completeness of nuclear final states $A^{\prime}$.
In the literature~\cite{Bauer:1977iq}, this cross section is also called the summed cross section.

The standard representation for $\Gamma_A$ in terms of the nucleon scattering amplitudes $\Gamma_N$ and $\Gamma_Y$  is, 
see, e.g.~\cite{Bauer:1977iq}:
\begin{equation}
\Gamma_A(\vec{b})=1-\prod_{k=1}^A \left(1-\Gamma_{N}(\vec{b}-\vec{s}_k)-\Gamma_{Y}(\vec{b}-\vec{s}_k)\right) \,, 
\label{eq:Gamma_A}
\end{equation}
 where $\vec{s}_k$ denotes the transverse coordinate of
$k$th nucleon in the nucleus. 
The nuclear amplitude $\Gamma_A(\vec{b})$ is an operator, whose first and second powers [see Eq.~(\ref{eq:cs1})]
are sandwiched between nuclear ground states. Therefore, only even powers of $\Gamma_{Y}$ contribute to the resulting 
cross section. Moreover, since powers of $\Gamma_{Y}^{\ast}(\vec{b^{\prime}}-\vec{s}_k) \Gamma_{Y}(\vec{b}-\vec{s}_k)$ 
in Eq.~(\ref{eq:cs1}) involve the same $k$th nucleon (proton-dissociation in $\Gamma_A^{\dagger}(\vec{b^{\prime}})\Gamma_A(\vec{b})$
takes place on the same nucleon), one does not need to take into account the effects of nucleon ordering and a non-zero longitudinal 
momentum transfer associated with the final state $Y$ 
in Eq.~(\ref{eq:Gamma_A}).

Substituting Eqs.~(\ref{eq:Gamma_N}) and (\ref{eq:Gamma_A}) in Eq.~(\ref{eq:cs1})
and assuming independent nucleon distributions, one obtains:
\begin{eqnarray}
&&\langle 0|\Gamma_A^{\dagger}(b^{\prime})\Gamma_A(b)|0 \rangle  - 
 \langle 0|\Gamma_A^{\dagger}(b^{\prime})|0 \rangle \langle 0|\Gamma_A(b)|0 \rangle 
\nonumber\\
&=& \left[\left(1-\frac{\langle \sigma \rangle}{2}T_A(b)-\frac{\langle \sigma \rangle}{2}T_A(b^{\prime})
+\frac{\langle \sigma \rangle^2}{16 \pi B_{\rm el}}T_A(b)
e^{-(\vec{b}^{\prime}-\vec{b})^2/(4B_{\rm el})}+\frac{\omega_{\sigma}\langle \sigma \rangle^2}{16 \pi^2 } T_A(b)
\int d^2 q_t^{\prime}\, e^{i \vec{q^{\prime}_t}(\vec{b}^{\prime}-\vec{b})} [f_{\rm pd}(t^{\prime})]^2
\right) \right]^A \nonumber\\ 
&-&  \left[\left(1-\frac{\langle \sigma \rangle}{2}T_A(b) \right)\right]^A
\left[\left(1-\frac{\langle \sigma \rangle}{2}T_A(b^{\prime}) \right)\right]^A 
\nonumber\\
 &\approx &
\left(\frac{\langle \sigma \rangle^2 }{16 \pi B_{\rm el}} e^{-(\vec{b}^{\prime}-\vec{b})^2/(4 B_{\rm el})} +
\frac{\omega_{\sigma}\langle \sigma \rangle^2}{16 \pi^2 }
\int d^2 q_t^{\prime} e^{i \vec{q^{\prime}_t}(\vec{b}^{\prime}-\vec{b})} [f_{\rm pd}(t^{\prime})]^2
 \right)
AT_A(b) e^{-\langle \sigma \rangle AT_A(b)} 
\,,
\label{eq:cs2}
\end{eqnarray}
where $T_A(b)=\int dz \rho_A(b,z)$ is the nuclear optical density with $\rho_A(b,z)$ being the nuclear distribution.
In the derivation of Eq.~(\ref{eq:cs2}), we used that both $B_{\rm el}$ and the effective slope of $f_{\rm pd}(t)$ are much smaller than the slope of 
the nuclear form factor (the effective nucleus radius) and, hence, the nuclear density can be evaluated at the impact parameter $\vec{b}$. 
  In the last line, 
 we exponentiated the powers of $A$, expanded the result in powers of 
the elastic and proton-dissociation cross sections, and
kept the first leading term. 
The terms neglected in the last line of Eq.~(\ref{eq:cs2}) contribute to the $t$-integrated 
$\sigma_{\gamma A \to J/\psi Y}(W_{\gamma p})$ cross section at the level of a few percent, which is well below the theoretical uncertainty associated with the 
calculation of the leading contribution to $\sigma_{\gamma A \to J/\psi Y}(W_{\gamma p})$ (see our numerical results in
Sec.~\ref{sec:upc}). 

\begin{figure}[ht!]
\begin{center}
\epsfig{file=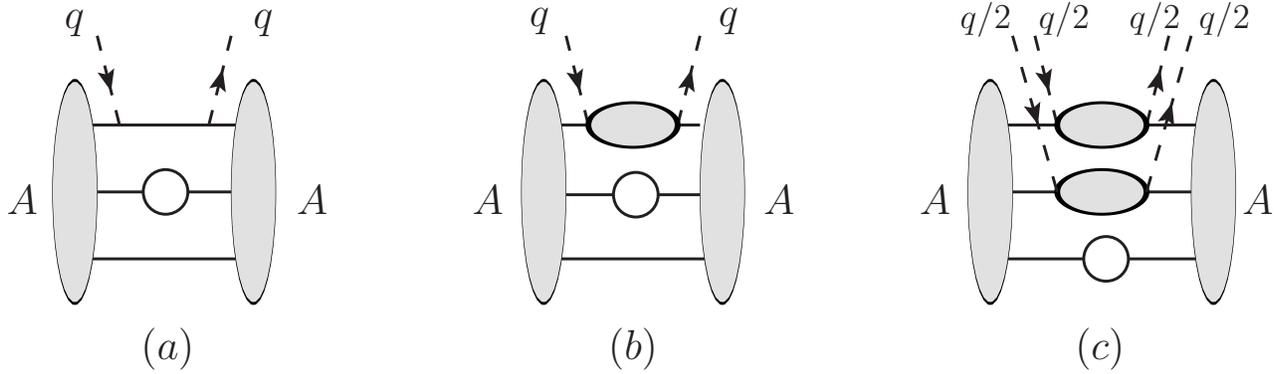,scale=1.1}
\caption{Schematic representation of incoherent $J/\psi$ photoproduction on nuclei: (a) contribution of elastic production on nucleons, (b) the proton-dissociation contribution, (c)  an example of terms neglected in the last line of Eq.~(\ref{eq:cs2}).
The dashed lines denote the momentum transfer $q$ to the interacting nucleon; the open circles on nucleon lines stand for
the interaction with that nucleon leading to attenuation of the resulting cross section; the vertical ovals labeled $A$ denote the nuclear states. }
\label{fig:inc_2018}
\end{center}
\end{figure}

In the graphic form, Eq.~(\ref{eq:cs2}) is schematically shown in Fig.~\ref{fig:inc_2018}, where graph $a$ denotes 
the contribution proportional to the elastic $J/\psi$ photoproproduction on the proton (nucleon), graph $b$ corresponds to
the proton-dissociation contribution
(the horizontal ovals denote nucleon dissociation), and graph $c$ is an example of terms proportional to higher powers of 
the proton-dissociation and elastic cross sections, which are neglected in the last line of Eq.~(\ref{eq:cs2}).
Note that these terms have a slower $t$-dependence because, as shown in the figure, they involve a fractional momentum
transfer.
The dashed lines denote the momentum transfer $q$ to the interacting nucleon; the open circles on nucleon lines stand for
the interaction with that nucleon, which leads to attenuation of the resulting incoherent cross section (all graphs with $0, 1, 2, \dots, A-1$ open circles contribute); the vertical ovals labeled $A$ denote the nuclear states. 

Substituting Eq.~(\ref{eq:cs2}) in Eq.~(\ref{eq:cs1}), we obtain
\begin{eqnarray}
\frac{d \sigma_{\gamma A \to J/\psi Y}(W_{\gamma p})}{dt}&=&\frac{\varkappa^2}{16 \pi} 
\left(\langle \sigma \rangle^2 e^{-q_t^2 B_{\rm el}}+\omega_{\sigma} \langle \sigma \rangle ^2 [f_{\rm pd}(t)]^2 \right)
 \int d^2 b
AT_A(b) e^{-\langle \sigma \rangle  AT_A(b)} \nonumber\\
&=&
\left(\frac{d\sigma_{\gamma p \to J/\psi p}(W_{\gamma p},t)}{dt}+\frac{d\sigma_{\gamma p \to J/\psi Y}(W_{\gamma p},t)}{dt}\right)
 \int d^2 b
AT_A(b) e^{-\langle \sigma \rangle  AT_A(b)}
\,,
\label{eq:cs3}
\end{eqnarray}
where in the last line we used Eq.~(\ref{eq:tdep}).
Equation~(\ref{eq:cs3}) has been derived using standard assumptions of the Gribov--Glauber model of nuclear shadowing and a generic representation of hadronic fluctuations of target nucleons. 
To include also the effect of hadronic fluctuations in the projective photon, we follow 
the procedure used in Ref.~\cite{Guzey:2013jaa} and express the eikonal factor in Eq.~(\ref{eq:cs3})
 in terms of the cross section $\sigma_2$, which is determined by the ratio of the diffractive and usual gluon densities in the proton, and $\sigma_3 \equiv \sigma_{\rm soft}$, which is modeled, see details in~\cite{Frankfurt:2011cs}.
It allows us to rewrite Eq.~(\ref{eq:cs3})
in the following final form:
\begin{equation}
\frac{d \sigma_{\gamma A \to J/\psi Y}(W_{\gamma p})}{dt}=\left(\frac{d\sigma_{\gamma p \to J/\psi p}(W_{\gamma p},t)}{dt}+\frac{d\sigma_{\gamma p \to J/\psi Y}(W_{\gamma p},t)}{dt}\right)  \int d^2 b AT_A(b) \left(1-\frac{\sigma_2}{\sigma_3}+ \frac{\sigma_2}{\sigma_3}e^{-\frac{\sigma_3}{2} AT_A(b)} \right)^2 
\,.
\label{eq:cs_final}
\end{equation}

Equation~(\ref{eq:cs_final}) generalizes Eq.~(15) of~\cite{Guzey:2013jaa} by including nucleon dissociation and has a clear physical interpretation (see Fig.~\ref{fig:inc_2018}): photoproduction of $J/\psi$ takes place on all $A$ target nucleons either elastically or with nucleon dissociation;
the interaction of photon hadronic fluctuations with remaining nucleons may lead inelastic production;
the probability not to 
have inelastic processes is given by the last term in the brackets in Eq.~(\ref{eq:cs_final}), which
describes the effect of nuclear shadowing.

\begin{figure}[h]
 \includegraphics[scale=1.0]{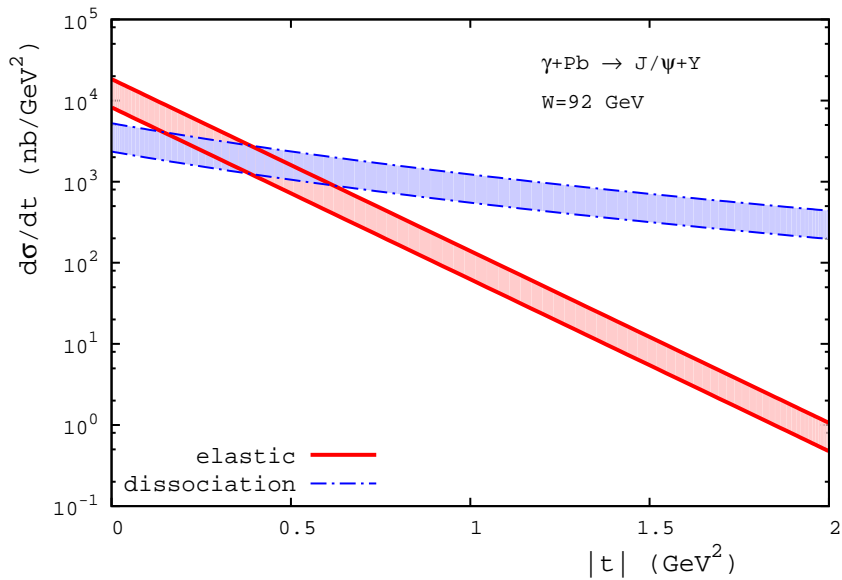}
 \caption{The elastic and nucleon-dissociation contributions to the $d \sigma_{\gamma A \to J/\psi Y}(W_{\gamma p})/dt$ cross section as a function of $|t|$ corresponding to the first and second terms in Eq.~(\ref{eq:cs_final}), respectively.
The shaded error bands quantify the uncertainty in the calculation of nuclear shadowing; 
$W_{\gamma p}=94$ GeV
corresponds to Pb-Pb UPCs at $\sqrt{s_{NN}}=2.76$ TeV and $y=0$. 
}
 \label{fig:incoh_tdep_nuc}
\end{figure}

Figure~\ref{fig:incoh_tdep_nuc} shows separately the $t$ dependence of the two contributions to the $d \sigma_{\gamma A \to J/\psi Y}(W_{\gamma p})/dt$ cross section in Eq.~(\ref{eq:cs_final}): the first term proportional to $d\sigma_{\gamma p \to J/\psi p}/dt$
is given by the red solid lines and the second term proportional to $d\sigma_{\gamma p \to J/\psi Y}/dt$ is given by the blue dot-dashed curves. The corresponding error bands reflect the uncertainty in the calculation of the nuclear shadowing effect, see the discussion in Sect.~\ref{sec:upc}; the value of the invariant photon--nucleon energy $W_{\gamma p}=94$ GeV
corresponds to the considered case of Pb-Pb UPCs at $\sqrt{s_{NN}}=2.76$ TeV and $y=0$.
This figure clearly demonstrates that while the elastic contribution dominates at $t \approx 0$, the nucleon-dissociation term
wins over for $|t| > 0.5$ GeV$^2$. After integration over $t$, the net contribution of the two terms is numerically close.

Several comments are in order here. 
First, the Good--Walker picture of hadronic fluctuations~(\ref{eq:fl1}) is valid strictly only at small $t \approx 0$. 
In our analysis, to apply completeness of final nuclear states $A^{\prime}$, we implied that $|t|$ is not small. However, there 
is no contradiction here: the momentum transfer $t$ is indeed small in all vertices except for the one leading to nucleon dissociation (see Fig.~\ref{fig:inc_2018}). For that vertex, we extrapolate to any $t$ using Eq.~(\ref{eq:tdep}), which allows us to write a simple expression
for $d \sigma_{\gamma A \to J/\psi Y}/dt$ valid for $t \neq 0$. Second, as follows from the second and third lines of Eq.~(\ref{eq:cs2}), the nuclear incoherent $d \sigma_{\gamma A \to J/\psi Y}/dt$ cross section in the impulse approximation should 
vanish in the $t=0$ limit due to a negative (coherent) contribution, which is concentrated at very small $t$; the shadowing correction makes the cross section finite. Since our final expression in Eq.~(\ref{eq:cs_final}) is applied at $|t| \neq 0$, 
this coherent correction is very small and can be safely neglected. 
Third, since the slopes of the $t$ dependence of $d\sigma_{\gamma p \to J/\psi p}/dt$ 
and $d\sigma_{\gamma p \to J/\psi Y}/dt$ are very different, the nucleon elastic and dissociation contributions
can be separated by studying the $t$ dependence of $d \sigma_{\gamma A \to J/\psi Y}(W_{\gamma p})/dt$.

For the $t$ integrated cross section, one readily obtains from Eq.~(\ref{eq:cs_final}):
\begin{equation}
\sigma_{\gamma A \to J/\psi Y}(W_{\gamma p}) =
\left(\sigma_{\gamma p \to J/\psi p}(W_{\gamma p})+\sigma_{\gamma p \to J/\psi Y}(W_{\gamma p})\right) \int d^2 b AT_A(b) \left(1-\frac{\sigma_2}{\sigma_3}+\frac{\sigma_2}{\sigma_3} e^{-\frac{\sigma_3}{2} AT_A(b)}\right)^2 \,.
\label{eq:cs_final_tint}
\end{equation}
One can see from Eq.~(\ref{eq:cs_final_tint}) that the nucleon elastic and dissociation contributions enter with 
equal nuclear shadowing factors. This can be seen by comparing graphs $a$ and $b$ in Fig.~\ref{fig:inc_2018}.
This result is derived neglecting graph $c$ and other similar graphs, whose
net numerical contribution to the $t$-integrated  $\sigma_{\gamma A \to J/\psi Y}(W_{\gamma p})$ cross section is a few percent correction.

\section{Incoherent $J/\psi$ photoproduction in Pb-Pb UPCs and comparison to ALICE data}
\label{sec:upc}

The cross section of incoherent $J/\psi$ photoproduction in symmetric nucleus--nucleus UPCs reads~\cite{Baltz:2007kq}:
\begin{equation}
\frac{d\sigma_{AA \to J/\psi AY}(y)}{dy}=N_{\gamma/A}(y) \sigma_{\gamma A \to J/\psi Y}(y)+
N_{\gamma/A}(-y) \sigma_{\gamma A \to J/\psi Y}(-y) \,,
\label{eq:cs_upc}
\end{equation}
where $N_{\gamma/A}$ is the photon flux; $y$ is the rapidity of the produced vector meson $V$; $\sigma_{\gamma A \to J/\psi Y}(y)$ 
is the nuclear incoherent cross section integrated over $t$.
The presence of two terms with opposite rapidities in Eq.~(\ref{eq:cs_upc}) reflects the fact that each colliding ion can serve
as a source of photons and as a target. Interference between these two contributions is important only for very small values of
$|t|$ and, hence, has been neglected.
The photon flux $N_{\gamma/A}(y)$ produced by an ultrarelativistic ion in nucleus--nucleus UPCs
is calculated using the standard expressions taking into account the effects of 
the strong interaction suppression and the nuclear form factor, see, e.g.~\cite{Guzey:2013xba}.
In our analysis, the nuclear density and form factor are calculated using the Hartree--Fock--Skyrme model~\cite{Beiner:1974gc}.

Substituting Eq.~(\ref{eq:cs_final_tint}) in Eq.~(\ref{eq:cs_upc}), we obtain our prediction
for the incoherent cross section $d\sigma_{AA \to J/\psi AY}(y=0)/dy$ of Pb-Pb UPCs at $\sqrt{s_{NN}}=2.76$ TeV
at the central rapidity $y=0$. It is shown by the first three lines of Table~\ref{table:res}, which correspond to the 
proton elastic contribution, the proton-dissociation one, and their sum. Each of the values in the first two lines bear 
10\% uncertainties due to experimental errors of the respective proton cross sections; the values in the third line come
with the combined 15\% experimental uncertainty.
Within significant theoretical uncertainties, which we will discuss below,
our prediction for $d\sigma_{AA \to J/\psi AY}(y=0)/dy$ (third line)
agrees with the ALICE experimental value~\cite{Abbas:2013oua} given in the last line of Table~\ref{table:res}.

\begin{table}[h]
\caption{Incoherent cross section $d\sigma_{AA \to J/\psi AY}(y=0)/dy$ of Pb-Pb UPCs at $\sqrt{s_{NN}}=2.76$ TeV
and $y=0$.}
\begin{center}
\begin{tabular}{|c|c|}
\hline
$d\sigma_{AA \to J/\psi AY}(y=0)/dy$: elastic & $0.32-0.68$ mb \\
$d\sigma_{AA \to J/\psi AY}(y=0)/dy$: dissociation & $0.27-0.56$ mb \\
$d\sigma_{AA \to J/\psi AY}(y=0)/dy$: total & $0.59-1.24$ mb \\
\hline
Experiment~\cite{Abbas:2013oua} & $0.98^{+0.19}_{-0.17}$ (sta+sys) mb \\
\hline
\end{tabular}
\end{center}
\label{table:res}
\end{table}

In our numerical analysis, we used the H1 data~\cite{Alexa:2013xxa} on the $\sigma_{\gamma p \to J/\psi p}$ and $\sigma_{\gamma p \to J/\psi Y}$ cross sections, see the end of Sec.~\ref{sec:fluct}.
One should note that
$r=\sigma_{\gamma p \to J/\psi Y}(W_{\gamma p})/\sigma_{\gamma p \to J/\psi p}(W_{\gamma p})$ depends on 
the maximal interval of rapidity, which is allowed for the final inelastic state $Y$. Therefore, the cuts used in incoherent 
$J/\psi$ photoproduction in UPCs should be consistent with those in $\gamma p\to J/\psi Y$.

The large range of resulting predictions for $d\sigma_{AA \to J/\psi AY}(y=0)/dy$ corresponds to the theoretical uncertainty of the leading twist nuclear shadowing model~\cite{Frankfurt:2011cs}, whose largest part is associated with the uncertainty in the effective cross 
section $\sigma_3$. It reflects uncertainties in modeling of the interplay between hard and soft components of diffraction in deep
inelastic scattering (DIS). At the same time,
the parameter $\sigma_2$ is constrained much better; a small uncertainty in $\sigma_2$ is related to experimental
errors of QCD analyses of hard diffraction at HERA, see details in~\cite{Frankfurt:2011cs}.
While these uncertainties lead to approximately $10$\% ambiguity in the predicted gluon nuclear shadowing in heavy nuclei at $x \approx 10^{-3}$, 
they are much larger for hard inelastic diffraction in DIS~\cite{Frankfurt:2011cs}. 
As one can see from Table~\ref{table:res}, incoherent photoproduction of $J/\psi$ on nuclei is also very sensitive 
to the value of $\sigma_3$. 
Thus, further studies of the discussed process would improve predictions for inclusive diffraction in DIS on nuclei, 
which is one of key measurements at a future Electron-Ion Collider (EIC)~\cite{Accardi:2012qut}.

To appreciate the magnitude of the leading twist nuclear shadowing suppression, one can cast our results in form of the following
ratio:
\begin{equation}
R=\frac{\sigma_{\gamma A \to J/\psi Y}(W_{\gamma p})}{A\left(\sigma_{\gamma p \to J/\psi p}(W_{\gamma p})+\sigma_{\gamma p \to J/\psi Y}(W_{\gamma p})\right)}=\int d^2 b T_A(b) \left(1-\frac{\sigma_2}{\sigma_3}+\frac{\sigma_2}{\sigma_3} e^{-\frac{\sigma_3}{2} AT_A(b)}\right)^2 \,.
\end{equation}
In the considered 
kinematics ($\sqrt{s_{NN}}=2.76$ TeV and $y=0$), we obtain
\begin{equation}
R=0.13-0.29 \,,
\label{eq:ration_num}
\end{equation}
which should be compared to unity in the limit of absence of nuclear shadowing. One can see from Eq.~(\ref{eq:ration_num})
that the effect of nuclear suppression due to the leading twist nuclear shadowing is even stronger than that in the case of 
coherent $J/\psi$ photoproduction in Pb-Pb UPCs~\cite{Guzey:2013xba,Guzey:2013qza}.

It was discussed in the literature that at large $|t| > 2$ GeV$^2$, the cross section of 
proton-dissociation $J/\psi$ photoproduction is proportional to the target gluon density $g_A(\tilde{x},|t|)$,
where $\tilde{x}= -t/(-t +M_Y^2-m_N^2)$~\cite{Frankfurt:2008et}. 
Hence, in a wide
range of $M_Y^2$ corresponding to $\tilde{x} \ge 10^{-2}$, where the effect of nuclear shadowing is weak, 
one would observe a nearly 
linear dependence of the cross section on $A$, which is much 
stronger than that given by Eq.~(\ref{eq:cs_final}) for small $|t|$.
At the same time, when $\tilde{x}$ is small, e.g., $\tilde{x} \sim 10^{-3}$,
the gluon nuclear shadowing slows down the $A$ dependence of the nuclear cross section.
In our approach, $g_A(x,\mu^2)$ can be readily evaluated in terms of $\sigma_2$ and $\sigma_3$~\cite{Guzey:2013jaa}:
\begin{equation}
g_A(x,\mu^2)=A g_N(x,\mu^2) \left[1-\frac{\sigma_2}{\sigma_3} +\frac{2\sigma_2}{A\sigma_3^2} \int d^2 b 
\left(1-e^{-AT_A(b) \frac{\sigma_3}{2}}\right)\right] \,,
\label{eq:gA}
\end{equation} 
where $g_N(x,\mu^2)$ is the gluon density in the nucleon. 
Despite the leading twist nuclear shadowing effect, the $A$ dependence of $g_A(x,\mu^2)$ encoded in Eq.~(\ref{eq:gA})
is still much faster than that given by Eq.~(\ref{eq:cs_final}). 
Indeed, transition to the dominance of the perturbative mechanism should result in a substantial increase of 
$R(t) \equiv d \sigma_{\gamma A \to J/\psi Y}(W_{\gamma p})/dt/(d\sigma_{\gamma p \to J/\psi p}(W_{\gamma p},t)/dt+d\sigma_{\gamma p \to J/\psi Y}(W_{\gamma p},t)/dt)$
 with an increase of $|t|$: a factor of $\ge 2$ in the discussed kinematics since 
 $g_A(\tilde{x},|t|)/[A g_N(\tilde{x},|t|)] \ge 0.6$ due to the leading twist nuclear shadowing.
Therefore, by studying the $A$ dependence of the  
$d \sigma_{\gamma A \to J/\psi Y}(W_{\gamma p})/dt$ cross section one should in principle distinguish between the
small and large $|t|$ regimes described by Eqs.~(\ref{eq:cs_final}) and (\ref{eq:gA}), respectively.

\section{Conclusions}
\label{sec:conclusion}

Using the general notion of cross section fluctuations in hadron--nucleus scattering at high energies, we derive an expression for 
the cross section of incoherent $J/\psi$ photoproduction on heavy nuclei $d\sigma_{\gamma A \to J/\psi Y}/dt$, which includes both elastic and proton-dissociation processes on target nucleons. The final expression for $d\sigma_{\gamma A \to J/\psi Y}/dt$ 
is given in terms of the sum of the 
$d\sigma_{\gamma p \to J/\psi p}/dt$ and $d\sigma_{\gamma p \to J/\psi Y}/dt$ cross sections times 
the common nuclear shadowing (suppression) factor. Using the HERA data for $\sigma_{\gamma p \to J/\psi p}$ and 
$\sigma_{\gamma p \to J/\psi Y}$
and the results of the leading twist nuclear shadowing model for the suppression factor, we made predictions 
for the cross section of incoherent $J/\psi$ photoproduction in Pb-Pb UPCs at $\sqrt{s_{NN}}=2.76$ TeV
and $y=0$, $d\sigma_{AA \to J/\psi AY}(y=0)/dy=0.59-1.24$ mb. Within large theoretical uncertainties of the leading twist nuclear
shadowing model for this cross section, our result agrees with the ALICE data point. The agreement is made possible by 
the large contribution to the nuclear incoherent cross section of the proton-dissociation process $\gamma p \to J/\psi Y$.
Thus, predictions of the leading twist nuclear shadowing model provide good description of both coherent and incoherent
$J/\psi$ photoproduction in Pb-Pb UPCs at the LHC.

\acknowledgments

VG would like to thank H.~M\"antysaari for useful discussions of the results of Refs.~\cite{Mantysaari:2016ykx,Mantysaari:2016jaz}.
The research of
MS  was supported  by the U.S. Department of Energy,
Office of Science, Office of Nuclear Physics, under Award No. DE-FG0 2-93ER40771.

\end{document}